\documentclass[nofootinbib,amsfonts,prd,aps]{revtex4}
\usepackage{graphicx}
\begin{document}

\newcommand{\cev}[1]{\reflectbox{\ensuremath{\vec{\reflectbox{\ensuremath{#1}}}}}}

\title{The Montevideo Interpretation of Quantum Mechanics: a short review}

\author{Rodolfo Gambini$^{1}$, Jorge Pullin$^{2}$}
\affiliation {
1. Instituto de F\'{\i}sica, Facultad de Ciencias, 
Igu\'a 4225, esq. Mataojo, 11400 Montevideo, Uruguay. \\
2. Department of Physics and Astronomy, Louisiana State University,
Baton Rouge, LA 70803-4001}

\begin{abstract}
  The Montevideo interpretation of quantum mechanics, which consists
  in supplementing environmental decoherence with fundamental
  limitations in measurement stemming from gravity, has been described
  in several publications. However, some of them  appeared
  before the full picture provided by the interpretation was
  developed. As such it can be difficult to get a good understanding via
  the published literature. Here we summarize it in a self contained
  brief presentation including all its principal elements.
\end{abstract}
\maketitle

\section{Introduction: the measurement problem}

Although quantum mechanics is a well defined theory in terms of
providing unambiguous experimental predictions that can be tested,
several physicists and philosophers of science find its presentation
to be unsatisfactory. At the center of the controversy is the well
known measurement problem. In the quantum theory states evolve
unitarily, unless a measurement takes place. During a measurement, the
state suffers a reduction that is not described by a unitary operator. In traditional
formulations, this non-unitary evolution is postulated. Such an
approach makes the theory complete from a calculational point of
view. However, one is left with an odd formulation: a theory that
claims our world is quantum in nature yet its own definition requires
referring to a classical world, as measurements are supposed to take
place when the system under study interacts with a classical
measurement device. 

More recently, a more careful inspection of how the interaction with a
measurement device takes place has led to a potential solution to the
problem. In the decoherence program (for a review and references see
\cite{decoherence}), the interaction with a measurement device and,
more generally, an environment with a large number of degrees of
freedom, leads the quantum system to behave almost as if a reduction
had taken place. Essentially the large number of degrees of freedom of
the measurement device and environment ``smother'' the quantum
behavior of the system under study. The evolution of the combined
system plus measurement device plus environment is unitary and
everything is ruled by quantum mechanics. But if one concentrates on
the wavefunction of the system under study only, tracing out the
environmental degrees of freedom, the evolution appears
to be non-unitary and very close to a reduction.

The decoherence program ---suitably supplemented by an ontology like
the Many Worlds one--- has not convinced everyone (see for instance
\cite{despagnat},\cite{ghirardi}) that it provides a complete solution to the
measurement problem. Objections can be summarized in two main points:

1) Since the evolution of the system plus environment plus measuring
device is unitary, it could happen that the quantum coherence of the
system being studied could be recovered. Model calculations show that
such ``revivals'' could happen, but they would take a long time for
most realistic measuring devices. However, it is therefore clear that
the picture that emerges is slightly different from the traditional
formulation where one can never dial back a reduction. A possible
answer is that for most real experimental situations one would have to
wait longer than the age of the universe. Related to this is the point
of when exactly does the measurement take place? Since all quantum
states throughout the evolution are unitarily equivalent, what
distinguishes the moment when the measurement takes place? Some have
put this as: ``in this picture nothing ever happens''. A possible
response is that after a certain amount of time the state of the
system is indistinguishable from the result of a reduction ``for all
practical purposes'' (FAPP) \cite{fapp}. But from a conceptual point
of view, a formulation of a theory should not rely on practical
aspects. One could imagine that future scientists could perhaps find
more accurate ways of measuring things and be able to distinguish what
today is ``FAPP'' indistinguishable from a reduction.

A related point is that one can define global observables for the
system plus measuring device plus environment \cite{despagnat,
  despagnat2}. The expectation value for one of these observables
takes different values if a collapse takes place or not. That could
allow in principle to distinguish the FAPP picture of decoherence from
a real collapse. From the FAPP perspective the answer is that these
types of observables are very difficult to measure, since it requires
measuring the many degrees of freedom of the environment. However, the
mere possibility of measuring these observables is not consistent with
a realistic description. This point has recently been highlighted by
Frauchiger and Renner, \cite{frauren} who show that quantum mechanics
is inconsistent with single world interpretations.

2) The ``and/or'' problem \cite{bell}. Even though the interaction
with the environment creates a reduced density matrix for the system
that has an approximate diagonal form, as all quantum states the
density matrix still represents a superposition of coexisting
alternatives. Why is one to interpret it as exclusive alternatives
with given probabilities? When is one to transition from an improper
to a proper mixture, in d'Espagnat's terminology \cite{despagnat}.

The Montevideo interpretation \cite{montevideo} seeks to address these
two criticisms. In the spirit of the decoherence program, it examines
more finely what is happening in a measurement and how the theory is
being formulated. It also brings into play the role of gravity in
physics. It may be surprising that gravity has something to do with
the formulation of quantum mechanics as one can imagine many systems
where quantum effects are very important but gravity seems to play no
role. But if one believes in the unity of physics it should not be
surprising that at some level one needs to include all of physics to
make certain situations work. More importantly, gravity brings to bear
on physics important limitations on what can be done. Non
gravitational physics allows to consider in principle arbitrarily
large amounts of energy in a confined region, which is clearly not
feasible physically if one includes gravity. This in particular places
limitations on the accuracy with which we can measure any physical
quantity \cite{salecker}.  Gravity also imposes limitations on our
notions of space and time, which are absolute in non-gravitational
physics. In particular one has to construct measurements of space and
time using real physical (and in this context, really quantum)
objects, as no externally defined space-time is pre-existent. This
forces subtle changes in how theories are formulated. In particular,
unitary theories do not appear to behave entirely unitarily since the
notion of unitary evolution is defined with respect to a perfect
classical time that cannot be approximated with arbitrary accuracy by
a real (quantum) clock \cite{obregon,torterolo}. Notice that the role
of gravity in this approach is different than in Penrose's
\cite{penrose}. Here the emphasis is on limitations to clocks due to
the intrinsically relational nature of time in gravity whereas in
Penrose's differences in time in different places is what is the basis
of the mechanism. 

These two new elements that the consideration of gravity brings to
bear on physics will be key in addressing the two objections to
decoherence that we outlined above. Since the evolution of systems is
not perfectly unitary, it will not help to revive coherence in quantum
systems to wait. Far from seeing coherence restored, it will be
progressively further lost. The limitations on measurement will impose
fundamental constraints on future physicists in developing means of
distinguishing the quantum states produced by decoherence from those
produced by a reduction. It will also make impossible to measure
global observables that may tell us if a reduction took place or
not. Notice that this is not FAPP: the limitations are fundamental. It
is the theories of physics that tell us that the states produced by
decoherence are indistinguishable from those produced by a
reduction. There is therefore a natural definition of when ``something
happens''. A measurement takes place when the state produced by
decoherence is indistinguishable from a reduction according to the
laws of physics \cite{axiomatic}. No invocation of an external observer is
needed. Measurements (more generally events) will be plentiful and
happening all the time around the universe as quantum systems interact
with the environment irrespectively of if an experimenter or measuring
device is present or not. The resulting quantum theory can therefore
be formulated on purely quantum terms, without invoking a classical
world. It also naturally leads to a new ontology consisting of quantum
systems, states and events, all objectively defined, in terms of which
to build the world. One could ask: weren't systems, states and events
already present in the Copenhagen interpretation? Couldn't we have
used them already to build the world?  Not entirely, since the
definition of event used there required the existence of a classical external
world to begin with. It therefore cannot be logically used to base the
world on.

In this small review we would like to outline some results supporting
the above point of view. In the next section we discuss how to use
real clocks to describe physical systems where no external time is
available. We will show that the evolution of the states presents a
fundamental loss of coherence. Notice that we are not modifying
quantum mechanics, just pointing out that we cannot entirely access
the underlying usual unitary theory when we describe it in terms of
real clocks (and measuring rods for space if one is studying quantum
field theories). In the following section we discuss how fundamental
limitations of measurement prevent us from distinguishing the state
produced by a reduction and a state produced by
decoherence. Obviously, given the complexities of the decoherence
process, we cannot show in general that this is the case. We will
present a modification of a model of decoherence presented by Zurek
\cite{zurek} to analyze this type of situation to exhibit the point we
are making. The next section discusses some philosophical implications
of having a realist interpretation of quantum mechanics like the one
proposed. We end with a summary.

\section{Quantum mechanics without an external time}

When one considers a system without external time, like when one
studies cosmology, or model systems like a set of particles with fixed
angular and linear momentum assuming no knowledge of external clocks
(see \cite{Anderson:2008sb} for references), one finds that the
Hamiltonian does not generate evolution but becomes a constraint that
can be written generically as $H=0$. One is left with what is called a
``frozen formalism'' (see \cite{anderson2,kuchar} and references
therein). The values of the canonical coordinates at a given time
$q(t),p(t)$ are not observable, since one does not have access to
$t$. Physical quantities have to have vanishing Poisson brackets with
the constraint, they are what is known as ``Dirac observables'' and
the canonical coordinates are not. The resulting picture is very
different from usual physics and it is difficult to extract physical
predictions from it since the observables are all constants of the
motion, as they have vanishing Poisson brackets with the
Hamiltonian. People have proposed several possible solutions to deal
with the situation although no general consensus on a solution
exists. We will not summarize all proposals here, in part because we
will not need most of them and for reasons of space. We will focus on
two proposals that, when combined, we claim  provide a satisfactory solution to how
to treat systems without external time when combined with each
other. For other approaches the review by Kucha\v{r} is very complete
\cite{kuchar}.

The first proposal we call ``evolving Dirac observables''. It has
appeared in various guises over the years, but it has been
emphasized by Rovelli \cite{rovelli}. The idea is to consider Dirac
observables that depend on a parameter $O(t)$. These are true Dirac
observables, they have vanishing Poisson brackets with the constraint
but their value is not well defined till one specifies the value of a
parameter. Notice that $t$ is just a parameter, it does not have to
have any connection with ``time''. The definition requires that when
the parameter takes the value of one of the canonical variables, the
Dirac observable takes the value of another canonical variable, for
example, $Q(t=q_1)=q_2$. This in part justifies why it is a Dirac
observable. Neither $q_1$ nor $q_2$ can be observed since we do not
have an external time, but the value $q_2$ takes when $q_1$ takes a
given value is a relation that can be defined without referring to an
external time, i.e. it is invariant information. As an example, let us
consider the relativistic particle in one dimension. We parameterize
it, including the energy as one of the canonical variables, $p_0$. One
then has a constraint $\phi=p_0^2-p^2 -m^2$. One can easily construct
two independent Dirac observables: $p$ and $X\equiv q-p\,
q^0/\sqrt{p^2+m^2}$ and verify they have vanishing Poisson brackets
with the constraint. An evolving constant of the motion could be,
\begin{equation}
  Q\left(t,q^a,p_a\right) = X+\frac{p}{\sqrt{p^2+m^2}}t,
\end{equation}
and one would have that when the parameter takes the value $q^0$, the
evolving constant $Q\left(t=q^0,q^a,p_a\right)=q$ takes the value of
one of the canonical variables. So one now has an evolution for the
system, the one in terms of the parameter $t$. But problems arise when
one tries to quantize things. There, variables like $q_1$ become
quantum operators but the parameter remains un-quantized. How does one
therefore make sense of $t=q_1$ at the quantum level when the left
member is a classical quantity and the right a quantum operator?
Particularly when the quantum operator is not a Dirac observable and
therefore not defined on the physical space of states of the theory.

The second approach was proposed by Page and Wootters
\cite{pagewootters}. They advocate quantizing systems without time by
promoting all canonical variables to quantum operators. Then one
chooses one as a ``clock'' and asks relational questions between the
other canonical variables and the clock.  Conditional probabilities
are well defined quantum mechanically. So without invoking a classical
external clock, one chooses a physical variable as a clock and to
study the evolution of probabilities one asks relational questions:
what is the expectation value of variable $q_2$ when variable $q_1$
(which we chose as clock) takes the value 3:30pm? Again, because
relational information does not require the use of external clocks, it
has invariant character and one can ask physical questions about
it. But trouble arises when one actually tries to compute the
conditional probabilities. Quantum probabilities require to compute
expectation values with quantum states. In these theories, since we
argued that the Hamiltonian is a constraint $H=0$, at a quantum level
one must have $\hat{H} \vert \Psi\rangle=0$, only states that are
annihilated by the constraint are permissible. But such space of
states is not invariant under multiplication by one of the canonical,
variables, i.e.  $\hat{H} q_1 \vert \Psi\rangle\neq 0$. So one cannot
compute the expectation values required to compute the conditional
probabilities. One can try to force a calculation pretending that one
remains in the space, but then one gets incorrect results. Studies of
model systems of a few particles have shown that one does not get the
right results for the propagators, for example \cite{kuchar}.

Our proposal \cite{torterolo} is to combine the two approaches we have
just outlined: one computes conditional probabilities of evolving
constants of the motion. So one chooses an evolving constant of the
motion that will be the ``clock'', $T(t)$, and then one chooses a
variable one wishes to study $O(t)$ and computes,
\begin{equation}
  P\left(O\in [O_0-\Delta_1 ,O_0+\Delta_1]\vert 
T\in [T_0-\Delta_2 ,T_0+\Delta_2]\right)=\lim_{\tau\to
\infty}\frac{\int_{-\tau}^\tau dt {\rm Tr}
\left(P_{O_0}^{\Delta_1}\left(t\right) P_{T_0}^{\Delta_2}\left(t\right)\rho
P_{T_0}^{\Delta_2}\left(t\right)\right)}{\int_{-\tau}^\tau dt {\rm
Tr}\left(P_{T_0}^{\Delta_2}\left(t\right)\rho\right)},
\end{equation}
where we are computing the conditional probability that the variable
$O$ takes a value within a range of width $2\Delta_1$ around the value
$O_0$ when the clock variable takes a value within a range of width
$2\Delta_2$ around the value $T_0$ (we are assuming the variables to
have continuous spectra, hence the need to ask about ranges of values)
on a quantum state described by the density matrix $\rho$. The
quantity $P_{O_0}^{\Delta_1}$ is the projector on the eigenspace
associated with the eigenvalue $O_0$ of the operator $\hat{O}$ and
similarly for $P_{T_0}^{\Delta_2}$. Notice that the expression does not
require assigning a value to the classical parameter $t$, since it is
integrated over all possible values.

We have shown \cite{torterolo} using a model system of two free
particles where we use one of them as ``clock'' that this expression,
provided one makes judicious assumptions about the clock, indeed
reproduces to leading order the correct usual propagator, not having
the problems of the Page and Wootters proposal.

The above expression in terms of conditional probabilities may look
unfamiliar. It is better to rewrite it in terms of an effective
density matrix. Then it looks exactly like the ordinary definition of
probability in quantum mechanics,
\begin{equation}
  P\left(O_0\vert T_0\right) = \frac{{\rm
      Tr}\left(P_{O_0}^{\Delta_1}\left(0\right) \rho_{\rm
        eff}\left(T_0\right)\right)}{{\rm Tr}\left(\rho_{\rm eff}\left(T_0\right)\right)},
\end{equation}
where on the left hand side we shortened the notation omitting mention
of the intervals, but they still are there. The effective density
matrix is defined as,
\begin{equation}
  \rho_{\rm eff}\left(T\right) =\int_{-\infty}^\infty dt
  U_s\left(t\right)\rho_s U_s^\dagger(t) {\cal P}_t\left(T\right),\label{effective}
\end{equation}
where we have assumed that the density matrix of the total system is a
direct product of that of the subsystem we use as clock $\rho_{\rm
  cl}$ and that of the subsystem under study $\rho_s$, and a similar
assumption holds for their evolution operators $U$. The probability,
\begin{equation}
  {\cal P}_t\left(T\right) = \frac{{\rm Tr}\left(P_{T_0}^{\Delta_2}(0) U_{\rm
        cl}\left(t\right) \rho_{\rm cl} U_{\rm
        cl}^\dagger\left(t\right)\right)}{\int_{-\infty}^\infty dt
    {\rm Tr}\left(P_{T_0}^{\Delta_2}\left(t\right)\rho_{\rm cl}\right)},
\end{equation}
is an unobservable quantity since it represents the probability that
the variable $\hat{T}$ take a given value when the unobservable
parameter is $t$. 

The introduction of the effective density matrix clearly illustrates
what happens when one describes ordinary quantum mechanics in terms of
a clock variable that is a real observable, not a classical 
parameter. Examining equation (\ref{effective}) we see in the right
hand side the ordinary density matrix evolving unitarily as a function
of the unobservable parameter $t$. If the probability ${\cal
  P}_t\left(T\right)$ were a Dirac delta, then the effective density
matrix would also evolve unitarily. That would mean that the real
clock variable is tracking the unobservable parameter $t$
perfectly. But no physical variable can do that, there will always be
a dispersion and the probability ${\cal P}_t\left(T\right)$ will have
non-vanishing support over a range of $T$. What this is telling us is
that the effective density matrix for the system at a time $T$ will correspond
to a superposition of density matrices at different values of the
unobservable parameter $t$. The resulting evolution is therefore
non-unitary. We see clearly the origin of the non-unitarity: the real
clock variable cannot keep track of the unitary evolution of quantum
mechanics. 

In fact if we assume that the clock variable tracks the unobservable
parameter almost perfectly by writing 
\begin{equation}
  {\cal P}_t\left(T\right) = \delta(t-T)+b(T) \delta"(t-T)+\ldots,
\end{equation}
(a term proportional to $\delta(t-T)'$ only adds an unobservable
shift), one can show that the evolution implied by (\ref{effective})
is generated by a modified Schr\"odinger equation,
\begin{equation}
  -i\hbar\frac{\partial \rho}{\partial T} = \left[\hat{H},\rho\right]+
\sigma(T) \left[\hat{H},\left[\hat{H},\rho\right]\right]+\ldots,
\end{equation}
where $\sigma(T)=d b(T)/dT$ is the rate of spread of the probability
${\cal P}_t\left(T\right)$ and $\rho=\rho_{\rm eff}(T)$. 

So we clearly see that when describing quantum mechanics in terms of a
real clock variable associated with a quantum observable rather than
with a classical parameter, the system loses unitarity and it is
progressively worse the longer one waits. 

The existence of the effect we are discussing is not controversial. In
fact one can make it as large as one wishes simply choosing a bad
clock. Bonifacio {\em et al.} \cite{bonifacio} have reinterpreted
certain experiments with Rabi oscillations as being described with an
inaccurate clock and indeed experimentally one sees the loss of
coherence described above. More recently it has been demonstrated with
entangled photons as well \cite{moreva}.

But the question still remains: can this effect be made arbitrarily
small by a choice of the clock variable? If one takes into account
gravity the answer is negative. Using non-gravitational quantum
physics Salecker and Wigner \cite{salecker} examined the question of
how accurate can a clock be. The answer is that the uncertainty in the
measurement of time is proportional to the square root of the length
of time one desires to measure and inversely proportional to the
square root of the mass of the clock. So to make a clock more accurate
one needs to make it more massive. But if one takes gravity into
account there clearly is a limitation as to how massive a clock can
be: at some point it turns into a black hole. Several phenomenological
models of this were proposed by various authors and they all agree
that the ultimate accuracy of a clock goes as some fractional power of
the time to be measured times a fractional power of Planck's time
\cite{models}. Different arguments lead to slightly different powers,
but the result is always that the longer one wishes to measure time
the more inaccurate the clocks become. For instance in the
phenomenological model of Ng and Van Dam \cite{models} one has that $\delta T \sim
T^{1/3} T_{\rm Planck}^{2/3}$. Substituting that in the modified
Schr\"odinger equation, its solution can be found in closed form, in
an energy eigenbasis,
\begin{equation}
\rho(T)_{nm} =\rho_{nm}(0) \exp\left(-i\omega_{nm}
  T\right)\exp\left(-\omega_{nm}^2 T^{4/3}_{\rm Planck} T^{2/3}\right),  
\end{equation}
where $\omega_{nm}$ is the Bohr frequency between the two energy
eigenstates $n$ and $m$. We see that the off diagonal terms of the
density matrix die off exponentially. Pure states evolve into mixed
states.

\section{Completing decoherence: the Montevideo interpretation}

\subsection{Decoherence with clocks based on physical variables} 

In this section we would like to analyze how the use of a physical
clock in the description of quantum mechanics we introduced in the
last section, combined with other limitations in measurement, will
help address the objections to environmental decoherence as a solution
to the measurement problem. We start by illustrating the idea of
decoherence (and the objections) using a of a well known
model of environmental decoherence due to Zurek \cite{zurek}, possibly
one of the simplest models one can consider that still captures the
complexities involved. 

\subsubsection{Zurek's model}
It
consists of a spin one half system representing the microscopic system
plus the measuring device, with a two dimensional Hilbert space $\{
\vert +\rangle,\vert -\rangle\}$. It interacts with an ``environment''
given by a bath of many similar two state ``atoms'' each with a two
dimensional Hilbert space $\{\vert +\rangle_k,\vert -\rangle_k\}$. If
there is no interaction with the environment the two spin states may
be taken to have
the same energy, we choose it to be zero, and all the atoms also are
chosen with zero energy. The interaction Hamiltonian is given by
\begin{equation}
H_{\rm int} = \hbar \sum_k \left( g_k \sigma_z \otimes \sigma_z^k
\otimes_{j\neq k} I_j\right).
\end{equation}
$\sigma_z$ is  a Pauli matrix acting on the state of the system.  It
has eigenvalues $+1$ for the spin eigenvector $\vert+\rangle$ and $-1$ for
$\vert-\rangle$. The operators $\sigma^k_z$ are similar but acting on the
state of the $k$-th atom. $I_j$ denotes the
identity matrix acting on atom $j$ and $g_k$ is the coupling constant.
It has dimensions of frequency and characterizes the coupling energy of
one of the spins $k$ with the system. 
The model can be thought physically as providing a representation 
of a photon propagating in a polarization analyzer.

Through the interaction, the initial state, which we can take as,
\begin{equation}
\vert\Psi(0)\rangle = \left(a\vert+\rangle + b\vert-\rangle\right) \prod_{k=1}^N \otimes \left[
\alpha_k\vert+\rangle_k +\beta_k \vert-\rangle_k \right],
\end{equation}
with $a,b,\alpha_k$ and $\beta_k$ complex constants, 
evolved using the Schr\"odinger equation becomes,
\begin{eqnarray}\label{3}
\vert\Psi(t)\rangle
&=& a \vert+\rangle \prod_{k=1}^N\otimes\left[
\alpha_k\exp\left(ig_k t\right)\vert+\rangle_k
+ \beta_k \exp\left(-ig_k t\right)\vert-\rangle_k\right]\\
&&+ b \vert-\rangle
\prod_{k=1}^N\otimes\left[
\alpha_k\exp\left(-ig_k t\right)\vert+\rangle_k
+ \beta_k \exp\left(ig_k t\right)\vert-\rangle_k\right].\nonumber
\end{eqnarray}
From it one can construct a density matrix for the system plus
environment and tracing out the environmental degrees of freedom one
gets a reduced density matrix for the system,
\begin{equation}
\rho_c(t) = \vert a\vert^2 \vert+\rangle\langle+\vert + \vert b\vert^2 \vert-\rangle\langle-\vert + z(t) ab^* \vert+\rangle\langle-\vert
+z^*(t) a^* b\vert-\rangle\langle+\vert,\label{12}
\end{equation}
where
\begin{equation}\label{5}
z(t) = \prod_{k=1}^N \left[\cos\left(2g_k t\right)+i\left(\vert\alpha_k\vert^2
-\vert\beta_k\vert^2\right) \sin\left(2 g_k t\right)\right].
\end{equation}

The complex valued function of time $z(t)$ determines the values of
the off-diagonal elements. If it vanishes the reduced density matrix
could be considered a ``proper mixture'' representing several outcomes
with their corresponding probabilities. 

We claim that with the modified evolution we discussed in the previous
section the usual objections to decoherence do not apply.  Recall
which are the usual objections: 

1) The quantum coherence is still
there.  Although a quantum system interacting with an environment with
many degrees of freedom will very likely give the appearance that the
initial quantum coherence of the system is lost, ---the density matrix
of the measurement device is almost diagonal---, the information about
the original superposition could be recovered for instance carrying
out a measurement that includes the environment.  The fact that such
measurements are hard to carry out in practice does not prevent the
issue from existing as a conceptual problem.

2) The ``and/or problem'': Since the density matrix has been obtained
by tracing over the environment, it represents an improper, not
proper, mixture: looking at equation (\ref{12}) there is no way to select
(even in some conceptual sense)  one of the components of the density
matrix versus the
others.

Let us discuss now the problem of revivals. In the model, the function
$z(t)$ does not die off asymptotically but is multi-periodic, after a
very long time the off-diagonal terms become large. Whatever
definiteness of the values of the preferred quantity we had won by the end of the
measurement interaction turns out in the very long run to have been
but a temporary victory. This is called the problem of revivals (or
``recurrence of coherence'', or ``recoherence''). This illustrates
that the quantum coherence persists, it was just transferred to the
environment and could be measured using global observables. 

\subsubsection{A more realistic model and real clocks}

To analyze the effects of limitations of measurement and the use of
real clocks in detail we will need to consider a more realistic model
of spinning particles \cite{luispe}, the previous model is too simple
to capture the effect of the use of real clocks. Although this model
is ``almost realistic'' it has the property that the system,
environment and measurement apparatus are all under control, as one
would need to measure a global observable, for instance. It consists
of a spin $S$ in a cavity with a magnetic field pointing in the $z$
direction.  A stream of $N$ ``environmental'' spins flows sideways
into the cavity and eventually exits it, the interactions last a
finite time determined by the time spent in the cavity. The flow of
particles that represents the environment is sufficiently diluted
such that we can ignore interactions among themselves. 

The interaction Hamiltonian for the $k$-th spin of
the environment is,
\begin{equation}
\hat{H}_k = \hat{H}^B_k + \hat{H}^{\rm int}_k,
\end{equation}
with,
\begin{equation}
\hat{H}^B_k = \gamma_1 B \hat{S}_z\otimes \hat{I} _k +\gamma_2
B\hat{I}\otimes S^{k}_z,
\end{equation}
and
\begin{equation}
\hat{H}^{\rm int}_k = f_k \left(
\hat{S}_x \hat{S}_x^k +
\hat{S}_y \hat{S}_y^k +
\hat{S}_z \hat{S}_z^k\right),
\end{equation}
where $f_k$ are the coupling constants between the spin and each of
the particles of the environment, $\gamma_1$ y $\gamma_2$ are the magnetic moments of the central
and environment spins respectively and the $\hat{S}$ are spin
operators.

For the complete system one can define an observable considered by d'Espagnat
\cite{despagnat}. It has the property that its expectation value is
different depending on if the state has suffered a quantum collapse or
not. It definition is,
\begin{equation}
\hat{M} \equiv \hat{S}_x \otimes\prod_k^N \hat{S}^k_x.
\end{equation}
One has that $\langle \hat{M} \rangle_{\rm collapse} = 0$ whereas,
\begin{equation}
\langle \psi\vert M\vert\psi\rangle =
ab^*\prod_k^N\left[\alpha_k\beta_k^* +\alpha^*_k\beta_k\right]
e^{-2i\Omega_k\tau} + a^*b
\prod_k^N\left[\alpha_k\beta_k^* +\alpha^*_k\beta_k\right]
e^{2i\Omega_k\tau} \ne 0,
\end{equation}
with $\Omega_k \equiv \sqrt{4f_k^2+B^2(\gamma_1-\gamma_2)^2}$ and
$\tau$ is the time of flight of the environmental spins through the chamber.
One can therefore determine experimentally if a collapse or not took
place measuring this observable.

However, if one considers the corrections to the evolution resulting
from the use of physical variables as clocks as we discussed in the
previous section, one has that,
\begin{eqnarray}
\langle \hat{M}\rangle &=&
a b^* e^{-i2 N \Omega T} e^{-4 N B^2 (\gamma_1-\gamma_2)^2 \theta}
\prod_{k}^N \left[ \alpha_k \beta_k^* e^{-16 B^2\gamma_1\gamma_2\theta}
+\alpha_k^* \beta_k\right]\nonumber\\
&&+b a^*
e^{i2 N \Omega T} e^{-4 N B^2 (\gamma_1-\gamma_2)^2 \theta}
\prod_{k}^N \left[ \alpha_k \beta_k^*
+\alpha_k^* \beta_ke^{-16 B^2\gamma_1\gamma_2\theta}\right],\label{19}
\end{eqnarray}
where $\Omega \equiv B(\gamma_1 -\gamma_2)$, $\theta \equiv
\frac{3}{2} T_{\rm P}^{4/3} \tau^{2/3}$, $\tau$ is the time of flight
of the environment spins within the chamber and $T$ is the length of
the experiment.

There exist a series of conditions for the experiment to be feasible
that imply certain inequalities,
\begin{eqnarray}
&&a) \qquad 1< f\tau =\frac{\mu\gamma_1 \gamma_2}{\hbar}\frac{\tau}{d^3},
\label{a}\label{20}\\
&&b) \qquad \Delta x \sim \sqrt{\frac{\hbar T}{m}}, \label{b}\\
&&c)\qquad f\ll \vert B(\gamma_1 -\gamma_2)\vert, \label{c}\\
&&d)\qquad \langle \hat{M} \rangle \sim
\exp\left(-6 N B^2(\gamma_1-\gamma_2)^2
T_{\rm Planck}^{4/3} \tau^{2/3}\right),\label{d}\label{24}
\end{eqnarray}
with $f$ the interaction energy between spins which was assumed
constant through the cell, $\mu$ the permeability of the vacuum, $d$
the impact parameter of the spins of the environment, $m$ their mass,
and $\Delta x$ the spatial extent of the environment particles.

Condition a) makes the coupling of the spins strong enough for 
 decoherence to occur; b) is to prevent the
particles of the environment from dispersing too much and therefore
making us unable to find them within the detectors at the end of the
experiment; c) is the condition for decoherence to be in the $z$
basis, as was mentioned; d) is an estimation of the
the expectation value of the observable when the effect of
the real clock is taken into
account. For details of the derivation of these conditions see
our previous paper \cite{undecidability}.

So the expectation value is exponentially damped and it becomes more
and more difficult to distinguish it from the vanishing value one has
in a collapse situation. A similar analysis allows to show that
revivals are also prevented by the modified evolution.  When the
multi-periodic functions in the coherences tend to take again the
original value after a Poincar\'e time of recurrence, the exponential
decay for sufficiently large systems completely hides the revival under
the noise amplitude.

Thus, the difficulties found in testing macroscopic superpositions in
a measurement process are enhanced by the corrections resulting of the
use of physical clocks.

\subsection{Why the solution is not FAPP}

But temporal decoherence involves exponentials, the troublesome terms
of decoherence become exponentially small but how does this
observation help to solve the problem of outcomes? In what follows we
will provide a criterion for the occurrence of events based on the
notion of undecidability.

When one takes into account the way that time enters in generally
covariant systems including the quantum fluctuations of the clock, the
evolution of the total system (system plus measurement apparatus plus
environment) becomes indistinguishable from the collapse. This is also
true for revivals and the observation of the coherences of the reduced
density matrix of the system plus the measuring device. We call such
situation ``undecidability''.  We are going to show that
undecidability is not only for all practical purposes (FAPP) but
fundamental.

From the previous discussion one can gather that as one considers
environments with a larger number of degrees of freedom and as longer
time measurements are considered, distinguishing between collapse and
unitary evolution becomes harder.  But is this enough to be a
fundamental claim?

Starting from (\ref{19}) and using the approximations (\ref{20}-\ref{24})
one can show \cite{undecidability} that,
\begin{equation} \langle \hat{M} \rangle \sim \exp\left(-6 N
B^2(\gamma_1-\gamma_2)^2 T_{\rm Planck}^{4/3} \tau^{2/3}\right) \equiv
e^{-K}.
\end{equation}
with
\begin{equation}
K \gg \frac{N^5 T_{\rm Planck}^{4/3} \hbar^{20/3}}{m^4
(\gamma_1 \gamma_2)^{8/3} \mu^{8/3}}.
\end{equation}
Is it possible to build a very large ensemble allowing to distinguish
this value from zero?

Brukner and Kofler \cite{koflerbrukner} have recently proposed that from
a very general quantum mechanical analysis together with bounds from
special and general relativity there is a fundamental uncertainty in
the measurements of angles even if one uses a measuring device of the
size of the observable Universe.
\begin{equation}
\Delta \theta \gtrsim \frac{l_P}{R},
\end{equation}
where $l_P \equiv \sqrt{\hbar G/c^3} \approx 10^{-35}m$.  If we take
the radius of the observable universe as a characteristic length, $R
\approx 10^{27}m$, we reach a fundamental bound on the measurement of
the angle,
\begin{equation}
\Delta \theta \ge 10^{-62}.
\end{equation}
To distinguish $\langle M \rangle$ from zero one needs to take into
account that the observable will have an error that depends on $\Delta
\theta$ (since for instance $\hat{S}_x$ and $\hat{S}_y$ will get
mixed) . If the error is larger than $\langle M \rangle$ there is no
way of distinguishing collapse from a unitary evolution {\em for
  fundamental, not practical reasons}. Therefore the solution is not
FAPP.

The expectation value of the observable is \cite{undecidability},
\begin{equation}
\langle \hat{M}^{\Delta \theta} \rangle \gtrsim e^{-K} \pm 
(\Delta \theta)^{2N} + \langle E(\Delta \theta) \rangle.
\end{equation}
with
\begin{equation}
K \gg \frac{N^5 T_{\rm Planck}^{4/3} \hbar^{20/3}}{m^4
(\gamma_1 \gamma_2)^{8/3} \mu^{8/3}}.
\end{equation}

Therefore for,
\begin{equation}
  N > \left(\frac{2 \ln\left(\frac{R}{\ell_P}\right)\left(m
        \left(\gamma_1 \gamma_2\right)^4\right)^{2/3}
      \mu^{8/3}}{T^{4/3} \hbar^{20/3}}\right)^{1/4} m
  \left(\gamma_1\gamma_2\right)^2 \sim 10^7, \label{30}
\end{equation}
it becomes undecidable whether collapse has occurred or not. That
means that no measurement of any quantity, even in principle, can
ascertain whether the evolution equation failed to hold.  Notice that
the above discussion was restricted to a given experiment Our present
knowledge of quantum gravity and the complexities of the decoherence
process in general does not allow us to prove
undecidability for an arbitrary experimental setup. Even models
slightly more elaborate than the one presented here can be quite
challenging to analyze. A different model, involving interaction of a
spin with bosons has also been analyzed with similar results
\cite{luispe2}.

This model exhibits the difficulties of trying to obtain generic
results concerning decoherence. 
Notice that expression (\ref{30}) depends on the magnetic moments of
the spins $\gamma_{1,2}$. If they were very large decoherence would
not take place. One would be in the presence of a macroscopic system
exhibiting quantum behavior. One does not expect such systems to exist
---at least in the terms described in the model---, but the model does
not rule them out.

\subsection{The problem of outcomes also known as the issue of
  macro-objectification}

The problem of macro-objectification of properties may be described
according with Ghirardi as follows: how, when, and under what
conditions do definite macroscopic properties emerge (in accordance
with our daily experience) for systems that, when all is said and
done, we have no good reasons for thinking they are fundamentally
different from the micro-systems from which they are composed?

We think that undecidability provides an answer to this problem.  We
will claim that events occur when a system suffers an interaction with
an environment such that one cannot distinguish by any physical means
if a unitary evolution or a reduction of the total system, including
the environment, took place. This provides a
criterion for the production of events, as we had anticipated. In
addition, we postulate (we call this the {\em ontological postulate}
in \cite{axiomatic}) that when an event occurs, the system chooses
randomly (constrained by their respective probability values) one of
the various possible outcomes. Having an objective criterion for the
production of events based on undecidability, answers the objections 
raised by  \cite{frauren} since the observer and the ``super observer'' now have
consistent descriptions.

Philosopher Jeremy Butterfield, who has written an assessment of the
Montevideo interpretation \cite{butterfield}, has observed that up to
now we have only provided precise examples of undecidability for
spinning particles. In that sense he considers that the fundamental
loss of coherence due to the use of quantum clocks and to the quantum
gravitational effects should be used in the context of a many world
interpretation because it helps to answer some of the long held
obstructions to the combination of the decoherence program with the
many worlds approach.

After a detailed analysis we do not believe that conclusion is inescapable.  Let
us assume the worse case scenario: that there are no further quantum
gravitational limitations for the measurements of other variables as
the ones obtained for the spin by Kofler and Bruckner (even though
there have been many proposals to alter uncertainty relations, see
references in \cite{nature}).  However, given
the fact that the distinction between a unitary evolution that
includes quantum time measurements or a quantum reduction  would
require an exponentially growing number of individual measurements in
order to have the required statistics for distinguishing a non
vanishing exponentially small mean value from zero. Limitations
referring to the existence of a finite number of physical resources in
a finite observable Universe would be enough to ensure undecidability.
Obviously further investigations are needed, but in a sense this is
the fate of all studies involving decoherence, it is just not possible
to develop  general proofs given the complexities involved. 

\section{Some philosophical implications}

If the fundamental nature of the world is quantum mechanical and we
adopt an interpretation that provides an objective criterion for the
occurrence of events, we are led to an ontology of objects and events.
The interpretation here considered makes reference to primitive
concepts like systems, states, events and the properties that
characterize them.  Although these concepts are not new and are
usually considered in quantum mechanics, one can assign them a
unambiguous meaning only if one has an interpretation of the theory.
For example, events could not be used as the basis of a realistic
ontology without a general criterion for the production of events that
is independent of measurements and observers.  

On the other hand, the concepts of state and system only acquire
ontological value when the events also have acquired it since they are
both defined through the production of events. Based on this ontology,
objects and events can be considered the building blocks of reality.
Objects will be represented in the quantum formalism by systems in
certain states and are characterized by their dispositions to produce
events under certain conditions.  In the new interpretation, events
are the actual entities.  Concrete reality accessible to our senses
is constituted by events localized in space-time.  As Whitehead
\cite{whitehead} recognized: ``the event is the ultimate unit of
natural occurrence.''  Events come with associated properties. Events
and properties in the quantum theory are represented by mathematical
entities called projectors. Quantum mechanics provides probabilities
for the occurrence of events and their properties.  When an event
happens, like in the case of the dot on the photographic plate in the
double-slit experiment, typically many properties are actualized. For
instance, the dot may be darker on one side than the other, or may
have one of many possible shapes.

Take for instance the hydrogen atom. It is a quantum system composed
by a proton and an electron. A particular hydrogen atom is a system in
a given state, it is an example of what we call an object and it has a
precise disposition to produce events.  Russell in The Analysis of
Matter \cite{russell}, asserts that ``the enduring thing or object of
common sense and the old physics must be interpreted as a world-line,
a causally related sequence of events, and that it is events and not
substances that we perceive.'' He thus distinguishes events as basic
particulars from objects as derived, constructed particulars.”  We
disagree with this point of view because it ignores the role of the
physical states. He adds: ``Bits of matter are not among the bricks out
of which the world is built. The bricks are events and bits of matter
are portions of the structure to which we find it convenient to give
separate attention.'' This is not the picture provided by 
quantum mechanics. An independent notion of object is
required: One can even have event-less objects in quantum
mechanics. For instance, when not measured, the hydrogen atom is an
object according to the definition above even though it is not
producing an event. The resulting ontology is such that objects and
events are independent concepts, they are not derived one from the
other. 

This is only a sketch of philosophical issues raised by the new
interpretation. We have a more complete discussion in \cite{hospitable}.

\section{Summary}

We have presented an easy to follow guide to the Montevideo
interpretation. Readers interested in an axiomatic formulation should
consult our previous paper \cite{axiomatic}. All the bibliography can
be found in \cite{montevideo}.

To summarize, the use of real physical variables to measure time implies a
modification in how one writes the equations of quantum mechanics. The
resulting picture has a fundamental mechanism for loss of
coherence. When environmental decoherence is supplemented with this
mechanism and taking into account fundamental limitations in
measurement, a picture emerges where there is an objective, observer
independent notion for when an event takes place. The resulting
interpretation of quantum mechanics, which we call the Montevideo
interpretation, is formulated entirely in terms of quantum concepts,
without the need to invoke a classical world. We have been able to
complete this picture for a simple realistic model of decoherence involving
spins. Studies of more elaborate models are needed to further corroborate the
picture. 

\section{Acknowledgements}
This work was supported in part by grant NSF-PHY-1305000, CCT-LSU,
Pedeciba and the Horace Hearne Jr. Institute for Theoretical Physics
at LSU.


\begin{references}

\bibitem{decoherence}M. Schlosshauer, ``Decoherence and the
  quantum-to-classical transition'' Springer, Heidelberg, Germany
  (2007); E. Joos, H. Zeh, C. Kiefer, D. Giulini, J. Kupsch and
  I.-O. Stamatescu, ``Decoherence and the appearance of a classical
  world in quantum theory (2nd Edition)''. Springer, Berlin, Germany
  (2003) See also http://www.decoherence.de

\bibitem{despagnat}
B. d'Espagnat ``Veiled reality'', Addison Wesley,
  New York (1995).

\bibitem{ghirardi}
G. Girardi, ``Sneaking a Look at God's Cards, Revised Edition:
Unraveling the Mysteries of Quantum Mechanics'', Princeton University
Press, Princeton, NJ (2007).


\bibitem{fapp} D. Wallace, ``Decoherence and ontology, or how I
  learned to stop worrying and love FAPP'' in  In
  S. Saunders. J. Barrett, A. Kent and D. Wallace (eds.), Many Worlds?
  Everett, Quantum Theory, and Reality, Oxford University Press,
  Oxford, UK (2010) [arXiv:1111.2189 [physics.hist-ph]]

\bibitem{despagnat2}
B. d'Espagnat, Found. Phys. {\bf 20}, 1147 (1990).

\bibitem{frauren} D. Frauchiger, R. Renner, ``Single-world
  interpretations of quantum theory cannot be self consistent'' 
arXiv:1604.07422. 

\bibitem{bell} J. S. Bell, ``Against measurement'' in ``Sixty two
  years of uncertainty'', A. I. Miller (editor), Plenum, New York,
  (1990). 


\bibitem{montevideo} The complete bibliography on the Montevideo
  Interpretation is {\tt in http://www.montevideointerpretation.org}

\bibitem{salecker}
E. Wigner, Rev. Mod. Phys. {\bf 29}, 255 (1957);
H. Salecker, E. Wigner, Phys. Rev. {\bf 109}, 571 (1958).

\bibitem{obregon}
  R.~Gambini, R.~Porto and J.~Pullin,
  Gen.\ Rel.\ Grav.\  {\bf 39}, 1143 (2007)
  [gr-qc/0603090].


\bibitem{torterolo}
  R.~Gambini, R.~A.~Porto, J.~Pullin and S.~Torterolo,
  Phys.\ Rev.\ D {\bf 79}, 041501 (2009)
  [arXiv:0809.4235 [gr-qc]].

\bibitem{penrose} R. Penrose ``The road to reality: A Complete Guide
  to the Laws of the Universe'' Vintage (2007). 

\bibitem{axiomatic}
  R.~Gambini, L.~P.~Garcia-Pintos and J.~Pullin,
  Stud.\ Hist.\ Philos.\ Mod.\ Phys.\  {\bf 42}, 256 (2011)
  [arXiv:1002.4209 [quant-ph]]

\bibitem{zurek} W. Zurek, Phys. Rev. {\bf D26} (1982) 1862. 

\bibitem{Anderson:2008sb} 
  E.~Anderson,
  Class.\ Quant.\ Grav.\  {\bf 26}, 135020 (2009)
  [arXiv:0809.1168 [gr-qc]].

\bibitem{anderson2}
 A.~Anderson,
  ``Thawing the frozen formalism: The Difference between observables
  and what we observe,'' in 
``Directions in general relativity, Volume 2'' B. L. Hu and T. A. Jacobson
(editors),  p 13-27 (2005).

\bibitem{kuchar} K. Kucha\v{r}, Int. J. Mod. Phys. {\bf D20}, 3 (2011).

\bibitem{rovelli} 
 C.~Rovelli,
  Phys.\ Rev.\ D {\bf 43}, 442 (1991).

\bibitem{pagewootters}
 D.~N.~Page and W.~K.~Wootters,
  Phys.\ Rev.\ D {\bf 27}, 2885 (1983).

\bibitem{bonifacio}
D. M. Meekhof, C. Monroe, B. E. King, W. M. Itano, and
D. J. Wineland, Phys. Rev. Lett {\bf 76}, 1796 (1996); M. Brune,
F. Schmidt-Kaler, A. Maali, J. Dreyer, E. Hagley, J. M. Raimond, and
S. Haroche, Phys. Rev. Lett. {\bf 76}, 1800 (1996); R. Bonifacio,
S. Olivares, P. Tombesi, D. Vitali Phys. Rev. {\bf A61}, 053802
(2000).

\bibitem{moreva}
  E.~Moreva, G.~Brida, M.~Gramegna, V.~Giovannetti, L.~Maccone and M.~Genovese,
  Phys.\ Rev.\ A {\bf 89}, no. 5, 052122 (2014)
  [arXiv:1310.4691 [quant-ph]].

\bibitem{models}
F. K\'arolhy\'azy, A. Frenkel, B. Luk\'acs in ``Quantum
concepts in space and time'' R. Penrose and C. Isham, editors,
Oxford University Press, Oxford (1986);  G.~Amelino-Camelia,
Mod.\ Phys.\ Lett.\ A {\bf 9}, 3415 (1994); Y.~J.~Ng and H.~van Dam,
Annals N.\ Y.\ Acad.\ Sci.\ {\bf 755}, 579 (1995)
[arXiv:hep-th/9406110]; Mod.\ Phys.\ Lett.\ A {\bf 9}, 335 (1994);
[arXiv:gr-qc/9603014]; S. Lloyd, J. Ng, Scientific American, November
(2004).

\bibitem{luispe}
  R.~Gambini, L.~P.~G.~Pintos and J.~Pullin,
  Found.\ Phys.\  {\bf 40}, 93 (2010)
  [arXiv:0905.4222 [quant-ph]].


\bibitem{undecidability}
  R.~Gambini, L.~P.~Garcia-Pintos and J.~Pullin,
  Int.\ J.\ Mod.\ Phys.\ D {\bf 20}, 909 (2011)
  [arXiv:1009.3817 [quant-ph]].

\bibitem{koflerbrukner} C. Brukner, J. Kofler, ``Are there fundamental
  limits for observing quantum phenomena from within quantum
  theory?'', [arXiv:1009.2654 [quan-ph]].

\bibitem{luispe2}
L.~P.~G.~Pintos, M. Sc. Thesis, Universidad de la Rep\'ublica,
Montevideo, Uruguay (2011).

\bibitem{butterfield}
  J.~Butterfield,
  Stud. Hist. Phil. Mod. Phys. 52, 75 (2015)
  [arXiv:1406.4351 [physics.hist-ph]].

\bibitem{nature}
  I.~Pikovski, M.~R.~Vanner, M.~Aspelmeyer, M.~S.~Kim and C.~Brukner,
  Nature Phys.\  {\bf 8}, 393 (2012)
  [arXiv:1111.1979 [quant-ph]].

\bibitem{whitehead}
A. N. Whitehead, ``Science and the modern world'' Free Press, New York
(1997). 

\bibitem{russell}
B. Russell, ``The analysis of matter'', Spokesman Books, Nottingham, UK (2007).

\bibitem{hospitable} R. Gambini, J. Pullin ``A hospitable universe:
  Addressing Ethical and Spiritual Concerns in Light of Recent
  Scientific Discoveries'' Imprint Academic (2018).

\end{references}
\end{document}